\documentclass[a4paper,11pt]{article}

\usepackage[left=2.5cm,right=2.5cm,top=2.5cm,bottom=2.5cm]{geometry}
\usepackage{graphicx}
\usepackage{amssymb}
\usepackage{multirow}
\usepackage{amsmath}
\usepackage{psfrag}
\usepackage{setspace}
\usepackage{subcaption}
\usepackage[colorlinks=true,bookmarks=true]{hyperref}
\usepackage{float}
\usepackage{tabularx}
\usepackage{array}
\usepackage{cite}
\usepackage[utf8]{inputenc}
\usepackage{authblk}

\hypersetup{citecolor=blue}

\begin{document}

\title{Retrolensing by a spherically symmetric naked singularity}
\author[1]{Gulmina Zaman Babar\thanks{gulminazamanbabar@yahoo.com}}
\author[2,3,4]{Farruh Atamurotov\thanks{atamurotov@yahoo.com}}
\author[5]{Abdullah Zaman Babar\thanks{abdullahzamanbabar@yahoo.com}}
\author[6]{Yen-Kheng Lim\thanks{yenkheng.lim@xmu.edu.my}}

\affil[1]{\normalsize{\textit{School of Natural Sciences, National University of Sciences and Technology, Sector H-12, Islamabad, Pakistan}}}
\affil[2]{\normalsize{\textit{Inha University in Tashkent, Ziyolilar 9, Tashkent 100170, Uzbekistan}}}
\affil[3]{\normalsize{\textit{Akfa University, Kichik Halqa Yuli Street 17,  Tashkent 100095, Uzbekistan}}}
\affil[4]{\normalsize{\textit{Department of Astronomy and Astrophysics, National University of Uzbekistan,Tashkent  100174, Uzbekistan}}}
\affil[5]{\normalsize{\textit{Department of Electrical Engineering, Air University, Islamabad, Pakistan}}}
\affil[6]{\normalsize{\textit{Department of Physics, Xiamen University Malaysia, 43900 Sepang, Malaysia}}}

\date{\normalsize{\today}}
\maketitle

\begin{abstract}
Considering a strong field limit, we investigate the retrolensing phenomenon in the vicinity of a Janis-Newman-Winicour (JNW) naked singularity embedded in a scalar field. We assume that the light rays from a nearby source are reflected by the photon sphere of the naked singularity, acting as a lens, to create a pair of images. The analytic expressions of the lensing coefficients $\bar{a}$ and $\bar{b}$ are obtained, which are scalar dependent and discordant with \cite{Bozza:2002b}. Moreover, considering the powerful supermassive black hole candidates Sgr A* and M87*,
we examined the influence of the scalar field on the apparent brightness and angular positions of the parity images. Our results are highlighted in correspondence with a non-scalar field gravity, specifically the Schwarzschild gravity. We have found that the brightness increases in the presence of the scalar field, while, the lensing coefficients $\bar{a}$, $\bar{b}$, the angular positions and the angular separation of the relativistic images experience a reverse effect.
\end{abstract}

\section{Introduction} \label{intro}
Black holes, a hidden reality of the Universe, have been the centre of curiosity for the past couple of decades.
Most recently, the Event Horizon Telescope (EHT) collaboration \cite{aki:2019a,aki:2019b,Akiyama:2019vo} made it possible to visualize the first ever image of the invisible gravitating cosmic body. However, the search has never stopped because these black holes, acting as strong gravitational lenses, provide an opportunity to discover the early stages and the exotic structure of the Universe. Schwarzschild black hole serves as the simplest gravitational lens model which in the first place was explored by Virbhadra and Ellis in the paper \cite{Virbha:2000a}. Later on, Bozza et~al. developed an analytical approach \cite{Bozza:2001a} to study the relativistic images of the spacetime. Holz and Wheeler considered the retro-Macho event by a Schwarzschild black hole in close proximately to the solar system \cite{Wheel:2002a}, an exhaustive analysis was carried out
to examine the retroimages in an ecliptic as well as non-ecliptic plane, and finally, explored that a
brighter image could be achieved for a perfectly aligned order of the lens system.
Subsequently, a remarkable contribution was made to investigate the process in terms of a logarithmic function for various spherically symmetric gravities \cite{Bozza:2002b,Bozza:2003a,Bozza:2004a,Bozza:2008a}, similarly, rotating Kerr black hole has been addressed in \cite{Bozza:2005a,Bozza:2006a}.
In Refs. \cite{Eiroa:2002b,Eiroa:2004a,Abu:2017a,Eiroa:2005a,Eiroa:2014a,Tsuk:2014a} the authors identified the Reissner-Nordstr$\mathrm{\ddot{o}}$m, Braneworld, charged scalar field and Tangherlini black holes as the most powerful lensing tools to explore the structure of galaxies, likewise details about Ellis wormhole can be found in the literature \cite{Tsuk:2016a,Tsuk:2017c,Tsuk:2017d}. Though lately, Tsukamoto et~al. successfully introduced a new variable different from the one employed in \cite{Bozza:2002b} to anticipate the retrolensing process in a Reissner-Nordstr$\mathrm{\ddot{o}}$m spacetime \cite{Tsuk:2017a,Tsuk:2017b}. Moreover, the authors pointed out that the error term  $O(b-b_c)$ of the deflection angle $\alpha(b)$ enunciated by Bozza basically reads $O(b-b_c)\log(b-b_c)$. The aforementioned technique shall also be implemented in our analysis by considering the supermassive black holes Sgr A* hosted by the centre of our galaxy and M87* in a nearby galaxy as the lens models.

Naked singularities are generally regarded as hypothetical relativistic theoretical objects without an event horizon, notwithstanding the fact, the authors have successfully shown that the final state of a gravitational collapse could lead to the formation of a bare singularity \cite{Eardley:1978tr,Christodoulou:1984mz,Joshi:2001xi,Joshi:2013dva,Crisford:2017zpi}. Based on various aspects of the gravitating bodies the Refs. \cite{Virbhadra:2002ju,Gyulchev:2008ff,Kovacs:2010xm,patail:2012b} have managed to draw a precise demarcation between black holes and naked singularities.
In this paper, our prime objective is to follow the notion of \cite{Tsuk:2017a} to recreate the retrolensing phenomenon in a Janis-Newman-Winicour
geometry derived by Janis et~al. \cite{Janis:1968zz}, which is a static and spherically symmetric solution to the Einstein-massless scalar field equations. The solution has also been found to coincide with previously existing metrics by Fisher and Wayman \cite{Fisher:1948yn,Wyman:1981bd}, however, the latter one was affirmed as JNW metric by Virbhadra in \cite{Virbhadra:1997ie}.
In common parlance, the JNW naked singularity is regarded as an extension of the Schwarzschild geometry coupled with a massless scalar field.
General lensing features of the spacetime are discussed in \cite{Vrbh:2008b} by splitting up the spacetime into three distinct categories: weakly, marginally and strongly naked singularities. So far, the JNW gravity is intently explored in the light of its specificities, viz: geodesic structure, accretion disk, ultra-high collision energies, geodesics in a magnetized vicinity, periodic orbits and shadow casting in the following articles \cite{Chowdhury:2011aa,Zhou:2014jja,patail:2012a,Babar:2016a,Babar:2017a,Josh:2019a,Josh:2020b,sub:2020a}.

The rest of our paper is organized as follows. In Sec.~(\ref{Retro}),
we work out the lensing coefficients in a strong field paradigm to probe the deflection angle of a light ray
in the Janis-Newman-Winicour spacetime. The light curves are used to study the brightness of the retroimages, additionally,
the angular position and the angular separation of the parity images are also examined in Sec.~(\ref{Magnification}).
In the end, Sec.~(\ref{con}) presents an overview report of all the investigations conducted in this paper.

\section{Retrolensing in a strong deflection limit in the JNW spacetime}\label{Retro}
The line element for the Janis-Newman-Winicour gravity derived from
the static and spherically symmetric Einstein-massless scalar field equations is defined as \cite{patail:2012a,Babar:2017a}
\begin{align}\label{Jnwmetric}
ds^2=-f^{\nu} dt^2+f^{-\nu}dr^2+r^2f^{1-\nu}(d\theta^2+\sin^2\theta\phi^2),
\end{align}
where $f$ along with the scalar field are
\begin{align}
f=\bigg(1-\frac{r_g}{r}\bigg) \quad \mathrm{and} \quad \Phi=\bigg(\frac{1-\nu^2}{2}\bigg)\ln f.
\end{align}
The solution admits two specific parameters $\nu$ and $r_g$, which are associated with the
ADM mass $M$ and scalar charge $q$ by the following expressions
\begin{align}\label{}
\nu=\frac{2M}{r_g}  \quad \mathrm{and} \quad r_g=2\sqrt{M^2+q^2}.
\end{align}
The values of $\nu$ lie within the range $(0,1)$ when $q\neq0$. The Schwarzschild metric is preserved by setting $\nu=1$ i.e. $q=0$ \cite{patail:2012a}.
Note that, increasing the scalar charge $q$ decreases the values of $\nu$.
The curvature singularity \cite{Kofinas:2015sna,Kofinas:2016lcz,Kofinas:2017gfv,Anagnostopoulos:2018jdq,Anagnostopoulos:2019mrc,Bonanno:2020qfu}
that also is the event horizon location for a Schwarzschild spacetime exists at $r=r_g$.

\begin{figure}[t]
\centering
   \includegraphics[scale=0.5]{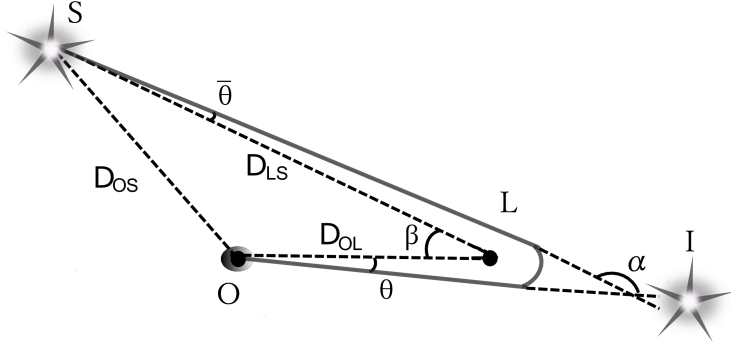}
\caption{Schematic representation of a retrolensing system typically based on the retrolens, source star and the image formed.}\label{Lensingsetup}
\end{figure}

\begin{figure*}[t]
 \begin{center}
   \includegraphics[scale=0.35]{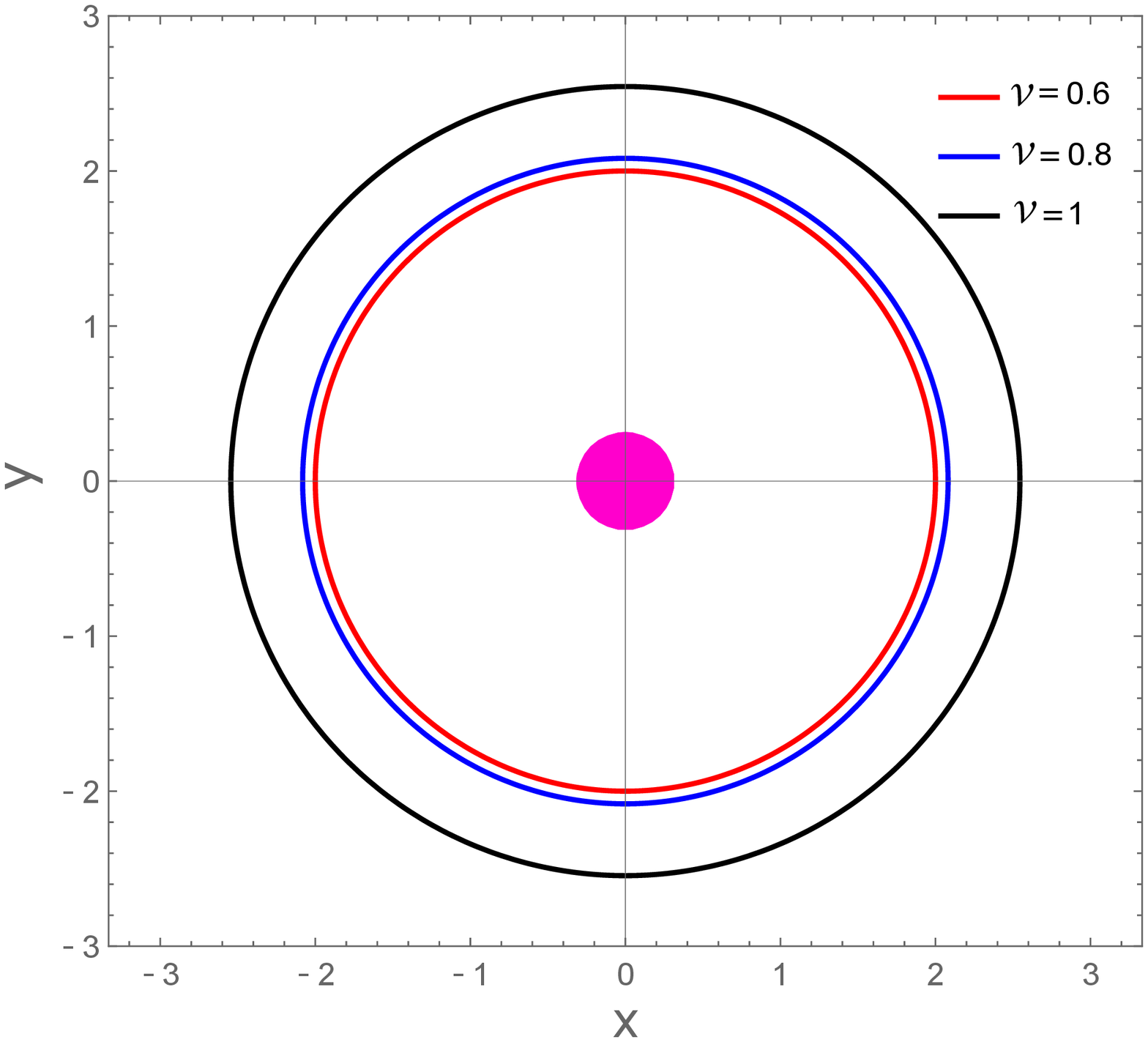}\hspace{1.5em}
    \includegraphics[scale=0.35]{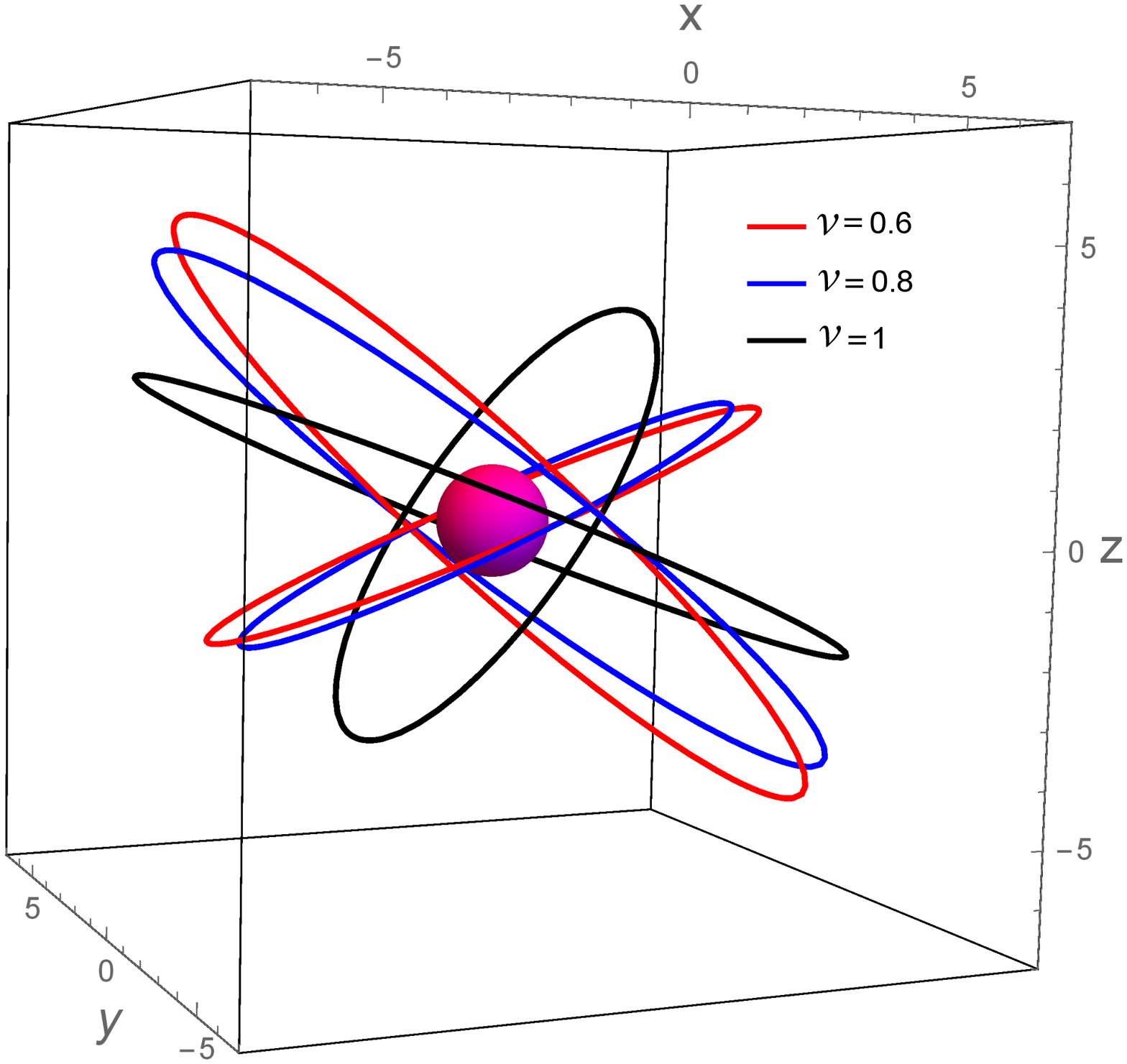}
  \end{center}
\caption{Left panel shows the light trajectories in a JNW geometry with fixed angular momentum $L$=5.5 and energy $E$=1. The red, blue and black plots represent $\nu$=0.6, 0.8 and 1, respectively. A 3D visualization is also shown in the right panel.}\label{orbits2d}
\end{figure*}

For the purpose of a meticulous analysis, we proceed with our investigation by considering the equatorial plane $\theta=\pi/2$. A customary gravitational lensing setup depending on the source star, retrolens and the image formed is illustrated in  Fig.~(\ref{Lensingsetup}). The light rays emitted from the source are reflected by the photon sphere
with a deflection angle $\alpha(b)$, generalized as
\footnote{In Ref \cite{Bozza:2002b}, the order of magnitude of the error term in the
deflection angle in the strong deflection limit reads as $O(b-b_c)$. Lately, on the contrary, with a high accuracy Ref \cite{Tsuk:2017a} identified the error term as $O((b-b_c)\log(b-b_c))$.} \cite{Tsuk:2017a}
\begin{align}\label{alfaone}
\alpha(b)=-\bar{a}\log\bigg(\frac{b}{b_c}-1\bigg)+\bar{b}+O((b-b_c)\log(b-b_c)),
\end{align}
here, $b$ and $b_c$ stand for the impact parameter and critical impact parameter of the light ray, respectively,
where, $b$ usually corresponds to the perpendicular distance between the flight path and the centre of a gravitating object.
The coefficients $\bar{a}$ and $\bar{b}$ are identified as constant quantities that are
evaluated at the radius of the photon sphere, $r_m$. One may easily compute $r_m$ as \cite{Tsuk:2017a}

\begin{align}
\frac{g^{'}_{\theta\theta}(r)}{g_{\theta\theta}(r)}- \frac{g^{'}_{tt}(r)}{g_{tt}(r)}=0,
\end{align}
which leads to

\begin{align}\label{photonrh}
r_m&= \frac{1}{2}(1+2\nu)r_g.
\end{align}
The stationary and spherically symmetric spacetime allows for two Killing vectors $\xi^{\mu}_{(t)}$ and $\xi^{\mu}_{(\phi)}$ that
result in the conserved quantities termed, respectively, as the energy $E$ and angular momentum $L$ of the massless particle \cite{patail:2012a}, given as

\begin{align}\label{engang}
E&=-g_{\mu\sigma}\xi^{\mu}_{(t)}\dot{x}^\sigma=f^{\nu}\dot{t}, \\
L&=g_{\mu\sigma}\xi^{\mu}_{(\phi)}\dot{x}^\sigma=r^2f^{1-\nu}\dot{\phi},
\end{align}
here, $\dot{x}^{\sigma}$ connotes the four velocity and overdot represents differentiation with respect to the affine parameter.
Further, these constants of motion are utilized to interpret the impact parameter $b$ as follows,
\begin{align}\label{impact}
b&=\frac{L}{E}=\frac{r^2f^{1-2\nu}\dot{\phi}}{\dot{t}}.
\end{align}
As established earlier, we study the motion along null geodesics in the equatorial plane, thus,
the flight path of a light ray defined by the normalization condition
$\dot{x}^{\sigma}\dot{x}_{\sigma}=0$ is obtained as

\begin{align}\label{metric2}
-f^{\nu}\dot{t}^2+f^{-\nu}\dot{r}^2+r^2f^{1-\nu}\dot{\phi}^2=0.
\end{align}
Equation (\ref{metric2}) is conventionally written in terms of the radial component and the effective potential as,

\begin{align}\label{Veff}
\dot{r}^2=E^2-V_\mathrm{eff} \quad \mathrm{where} \quad V_\mathrm{eff}=\frac{L^2}{r^2f^{1-2\nu}}.
\end{align}
Consistent with the inequality $r_m>r_g$ \cite{Babar:2017a}, the dynamics of photons is possible in the spacetime provided that $V_\mathrm{eff}>0$.
In Fig.~(\ref{orbits2d}), the photon orbits for distinct scalar parameter $\nu$ are plotted by numerically solving (\ref{Veff}) for fixed energy $E$=1 and angular momentum $L$=5.5. The location and inclination of the orbits manifestly alters by increasing the scalar field intensity. The trajectory (\ref{metric2}) can be recasted in the form
\begin{align}\label{lighttraj}
\bigg(\frac{\dot{r}}{\dot{\phi}}\bigg)^2=r^4f^{2(1-\nu)}\bigg(\frac{1}{b^2}-\frac{1}{r^2f^{1-2\nu}}\bigg).
\end{align}

\begin{figure}[t]
 \begin{center}
   \includegraphics[scale=0.42]{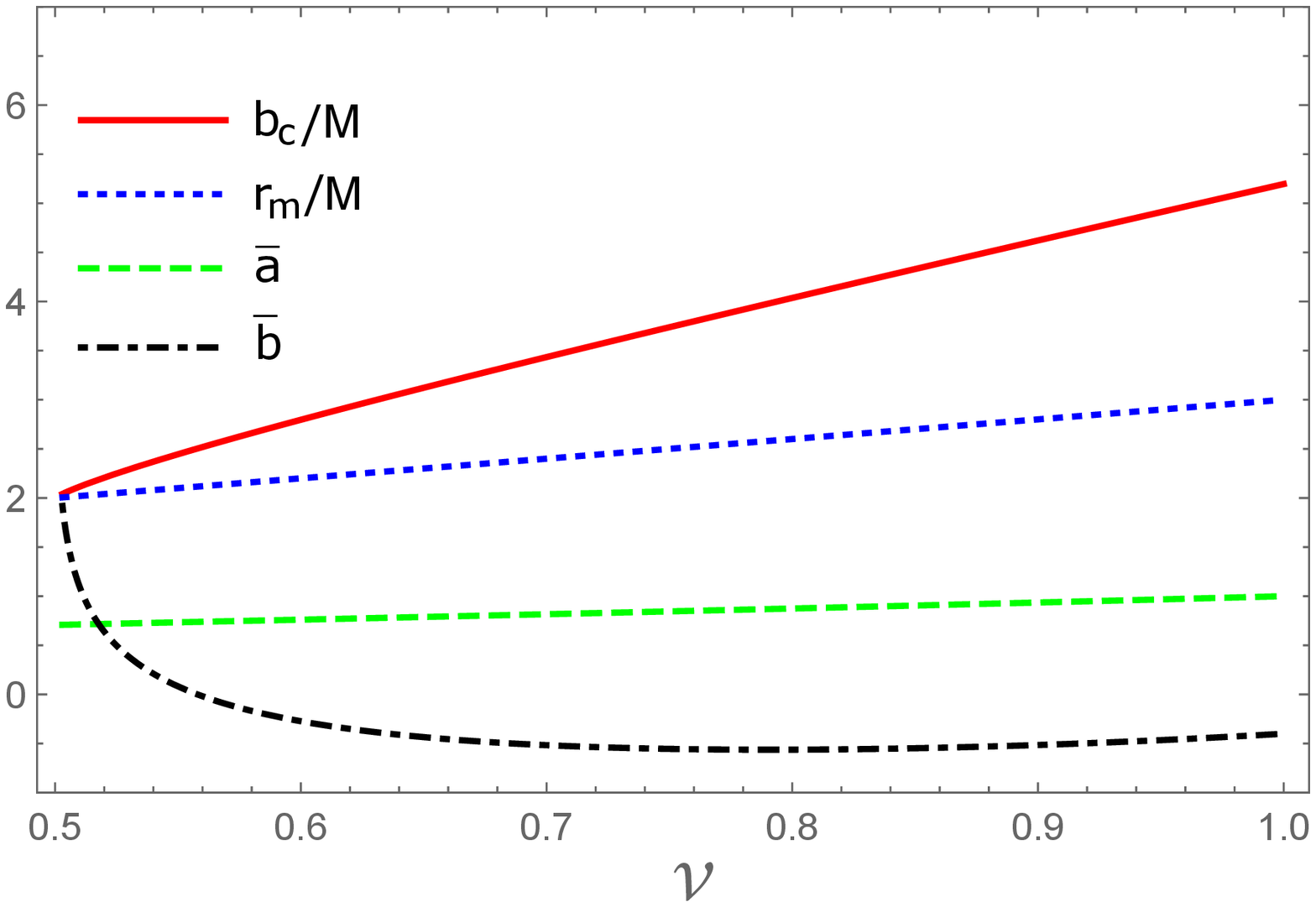}
   \includegraphics[scale=0.42]{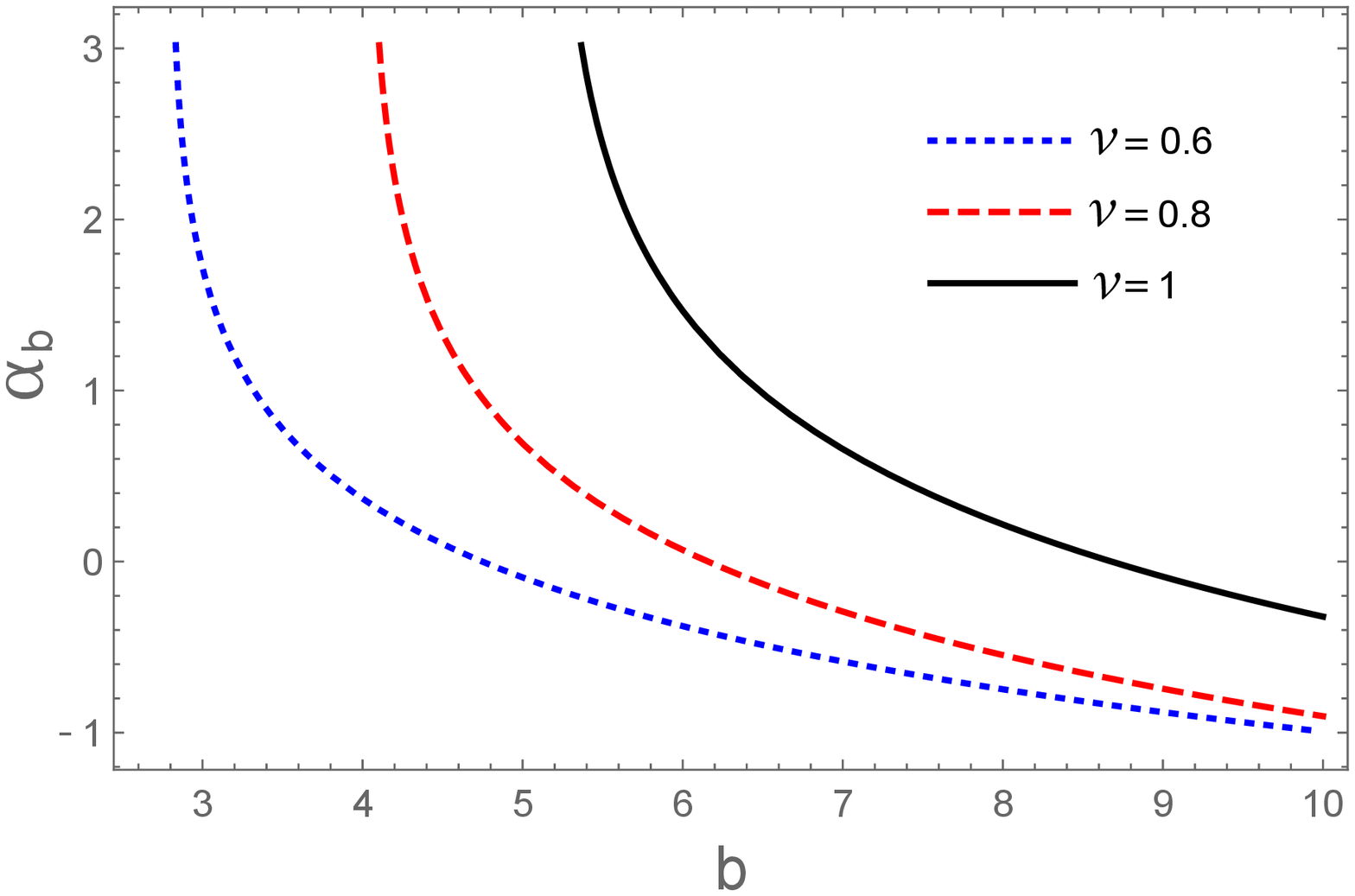}
  \end{center}
\caption{$b_c/M$ (red solid), $r_m/M $ (blue dotted), $\bar{a}$ (green dashed) and $\bar{b}$ (black dotdashed)
in the JNW spacetime as a function of $\nu$ (Left panel). Deflection angle $\alpha(b)$ as a function the impact parameter $b$
for different values of $\nu$ (Right panel).}\label{abrm}
\end{figure}

Now, to properly assess the strong field retrolensing in the JNW spacetime,
we establish a schema in a specific way for the photons engaged in the process.
Initially, the photons are considered at infinity and then allowed to fall freely towards the singularity
up to the range at the closest point $r=r_0$, where they are immediately reflected back to infinity.
As a special case, when $r_0 \to r_m$ we get the critical parameters as $b_{c}(r_m)=\lim_{r_0 \to r_m}b(r_0)$.
Using (\ref{Veff}), $b$ with respect to the point $r_0$ takes the form

\begin{align}
b(r_0)=r_0\sqrt{f^{1-2\nu}}.
\end{align}
Also, the deflection angle, $\alpha(r_0)$ \cite{Tsuk:2017a,Tsuk:2017b} is defined as
\begin{align}
\alpha=I(r_0)-\pi,
\end{align}
where $I(r_0)$ admits the expression
\begin{align}\label{Int}
I(r_0)=2\int_{r_{0}}^{\infty}\frac{1}{r^2f^{1-\nu}\sqrt{\bigg(\frac{1}{b^2}-\frac{1}{r^2f^{1-2\nu}}\bigg)}}dr.
\end{align}
Here, we shall employ the notion of \cite{Tsuk:2017a} to evaluate $\bar{a}$ and $\bar{b}$ by introducing a unique variable $z$ \footnote{In Ref \cite{Bozza:2002b}, a counterpart $z_{[1]}$=$\frac{-g_{tt}(r)+g_{tt}(r_0)}{1+g_{tt}(r_0)}$ of the variable $z$ is used with a numerical approach, however, Ref \cite{Tsuk:2017a} introduced $z$ defined by (\ref{ztup}) and successfully obtained an analytical expression of the deflection angle in the strong field limit of the Reissner-Nordstr$\mathrm{\ddot{o}}$m spacetime.} given by

\begin{align}
z=1-\frac{r_{0}}{r}.\label{ztup}
\end{align}
\\
By bringing in the new variable $z$ we ought to modify the limits $r_0$ and $\infty$, which respectively becomes 0 and 1.
Thus, (\ref{Int}) takes the form

\begin{align}\label{Int1a}
I(r_0)=2\int_{0}^{1}\sqrt{\frac{r_0}{c_1z+c_2z^2+c_3z^3}}dz,
\end{align}
where the order of divergence is $z^{-1}$ and the constant values are defined by

\begin{align}\label{constants}
  c_1(r_0)&=2r_0-(1+2\nu)r_g, \\
  c_2(r_0)&=-r_0+3r_g,  \\
  c_3(r_0)&=-r_g.
\end{align}
While dealing with the strong deflection limit, i.e, $r_0 \to r_m$, the constants $c_1(r_0)$ and $c_2(r_0)$ become,

\begin{align}\label{constants2}
c_1(r_0)&=0, \\
c_2(r_0)&= \frac{1}{2}(5-2\nu)r_g.
\end{align}
\\
Considering the strong field limit ($r_0 \to r_m$ or $b \to b_c$)
we resolve $I(r_0)$ distinctively for its divergent $I_D(r_0)$ and regular $I_R(r_0)$ part,
which is basically defined by $I(r_0)=I_D(r_0)+I_R(r_0)$ \cite{Tsuk:2017a,Tsuk:2017b}. The divergent part $I_D$ reads

\begin{align}\label{inta}
I_D=\lim_{r_0 \to r_m}2\int_{0}^{1}\sqrt{\frac{r_0}{c_1z+c_2z^2}}dz.
\end{align}
The solution of the above mentioned equation in terms of $\bar{a}$ is worked out by means of (\ref{photonrh})
in addition with the critical impact parameter $b(r_m)$
\begin{align}
I_D(b)&=-\bar{a}\log\bigg(\frac{b}{b_c}-1\bigg)+\bar{a}\log\bigg(\frac{4r_m}{(2\nu-1)r_g}\bigg)+O((b-b_c)\log(b-b_c)).
\end{align}
Surprisingly, unlike \cite{Bozza:2002b}, where $\bar{a}$ admits a constant value of unity,
we get a scalar dependent analytic expression for $\bar{a}$ as follows
\begin{align}\label{abar}
\bar{a}=\sqrt{\frac{2r_m}{r_g(5-2\nu)}}.
\end{align}
The regular part $I_R$ admits the following analytic expression
\begin{align}\label{Intb}
I_R(b)=\lim_{r_0 \to r_m}2\int_{0}^{1}\bigg(\sqrt{\frac{r_0}{c_2z^2+c_3z^3}}-\sqrt{\frac{r_0}{c_2z^2}}\bigg)dz.
\end{align}
Operating in the same way earlier upon (\ref{inta}), we substitute the respective critical parameters in
(\ref{Intb}) and attain the following result for $I_R$ in a strong field approximation

\begin{flalign}
I_R(b)=&\bar{a}\log\bigl(4(5-2\nu)^2\bigl(\sqrt{(3-2\nu)(5-2\nu)}-(4-2\nu)\bigr)^2\bigr)+O((b-b_c)\log(b-b_c)).
\end{flalign}
Finally, we find the value of $\bar{b}$ in the same manner as in \cite{Tsuk:2017a}

\begin{flalign}
\bar{b}=&\bar{a}\log\bigg(\frac{16r_m(5-2\nu)^2}{r_g(2\nu-1)}\big(\sqrt{(3-2\nu)(5-2\nu)}-(4-2\nu)\big)^2\bigg)-\pi.\label{bbar}
\end{flalign}
\\
The left panel of Fig.~(\ref{abrm}) displays $b_c/M$, $r_m/M $, $\bar{a}$ and $\bar{b}$ as a function of the scalar parameter $\nu$.
For $\nu=1$, we precisely reproduce the Schwarzschild results of \cite{Bozza:2002b}, i.e., $b_c=3\sqrt{3}M$, $r_m=3M$,
$\bar{a}=1$ and $\bar{b}=\log[216(7-4\sqrt{3})]-\pi$. The quantities
$b_c/M$, $r_m/M $ and $\bar{a}$ increase monotonically, whereas, $\bar{b}$ shows a decrease until a local minima is achieved
at $\approx 0.790825$. Note that, $\nu$=$\frac{1}{2}$ is the critical point
at which $b_c$, $r_m$ and $\bar{b}$ coincides with the curvature singularity, thereby, to follow null geodesics across the region delimited by
$\nu<$$\frac{1}{2}$  is not a suitable choice. Henceforward, in the upcoming analysis we shall accordingly restrict to the range $\frac{1}{2}\leq\nu\leq1$,
wherein, $\bar{a}>$0 and $\bar{b}<0$.
Table~(\ref{table1}) provides a comparison of the values of $\bar{a}$ and $\bar{b}$ in
a strong field limit with the values evaluated in \cite{Bozza:2002b}. The values of (\ref{abar}) are scalar dependent and
therefore do not yield a constant quantity as in \cite{Bozza:2002b}, also, (\ref{bbar}) holds a non-monotonic behaviour. Furthermore, the lensing coefficients are employed to explicate the deflection angle $\alpha(b)$ of the intervening massless particles, see the right panel,
one observes that the photons are strongly deviated near the naked singularity.

\begin{table*}[t]
\caption{The numerical values of $\bar{a}$ and $\bar{b}$ for different values of the scalar parameter $\nu$ are presented in contrast to \cite{Bozza:2002b}.
\label{table1}}
\resizebox{1\textwidth}{!}{
\begin{tabular}{p{2.6cm} p{2.6cm} p{2.8cm} p{2.6cm} p{2.6cm} p{2.6cm} p{2.6cm}}
\hline\hline
$\nu$                             & 1        & 0.9        & $0.8$      & 0.7         & 0.6 \\
\hline
$\bar{a}$                         &1.0000    &0.9354       &0.8745   &0.8165     &0.7609  \\
$\bar{a}_{\mathrm{Bozza}}$\cite{Bozza:2002b}        &1.0000   &1.0000        &1.0000   &1.0000     &1.0000        \\
$\bar{b}$                         &-0.4002  &-0.5153      &-0.5626   &-0.5172    &-0.2739      \\
$\bar{b}_{\mathrm{Bozza}}$\cite{Bozza:2002b}        &-0.4002  &-0.3808      &-0.3500   &-0.2945    &-0.1659            \\
\hline\hline
\end{tabular}}	
\end{table*}

\section{Lens system and magnification in a strong deflection limit}\label{Magnification}
This section attempts to discuss the magnification of retrolensed light beams in a JNW spacetime, to this end
a framework is required to apprehend the process, thus, we proceed with a setup motivated by the well-known Ohanian deflection system \cite{Ohanian:1987a,Tsuk:2017a}.

First of all, we briefly revisit the physical arrangement of the lens system discussed in \cite{Tsuk:2017a}. In Fig.~(\ref{Lensingsetup}), the distances from the lens to the source, from the lens to the observer and from the observer to the source are labeled as $D_{LS}$, $D_{OL}$ and $D_{OS}$, respectively.
Here, $S$ is the luminous source of light rays that are mirrored by the light sphere $L$ of the naked singularity and
as a result the observer $O$ views an image $I$ with an angle $\theta$.
In the Ohanian deflection system the source angle $\beta \in [0,\pi]$ is defined as \cite{Ohanian:1987a}

\begin{align}\label{beta1}
\beta=\pi-\alpha(\theta)+\theta+\bar{\theta},
\end{align}
\\
where, $\bar{\theta}$ is the angle between the light ray emitted from the source and the line $LS$.
The condition $\beta\sim0$ corresponds to a perfectly aligned case of the lens system where the naked singularity, the observer and the source are collinear. The value of $\theta_+(\beta)$ in a strong field limit derived in \cite{Tsuk:2017a} is

\begin{align}\label{theta+}
\theta_{+}=\theta_{m}\bigg(1+\mathrm{exp}\bigg(\frac{\bar{b}-\pi+\beta}{\bar{a}}\bigg)\bigg),
\end{align}
where $\theta_m=b_c/D_{OL}$ is the image angle of the photon sphere
of the naked singularity. In case of a negative solution we have $\theta_{-}(\beta)=-\theta_{+}(-\beta)\sim-\theta_{+}(\beta)$.
The total amplification of the relativistic image is defined as \cite{Bozza:2004a,Tsuk:2017a}
\begin{align}\label{mutot}
 \mu_{\mathrm{tot}}(\beta)&=2\frac{D_{OS}^2}{D_{LS}^2}\frac{\theta_m^2e^{\frac{\bar{b}-\pi}{\bar{a}}}(1+e^{\frac{\bar{b}-\pi}{\bar{a}}})}{\bar{a}}|s(\beta)|.
\end{align}
The of value $s(\beta)$ is evaluated in terms of the integral over a finite uniform-luminous disk placed in the source plane.
Remark that, our post-discussion follows a perfectly aligned case of the lens system.

\subsection{Light curves}\label{lightcurves}
\begin{figure}[t]
 \begin{center}
   \includegraphics[scale=0.269]{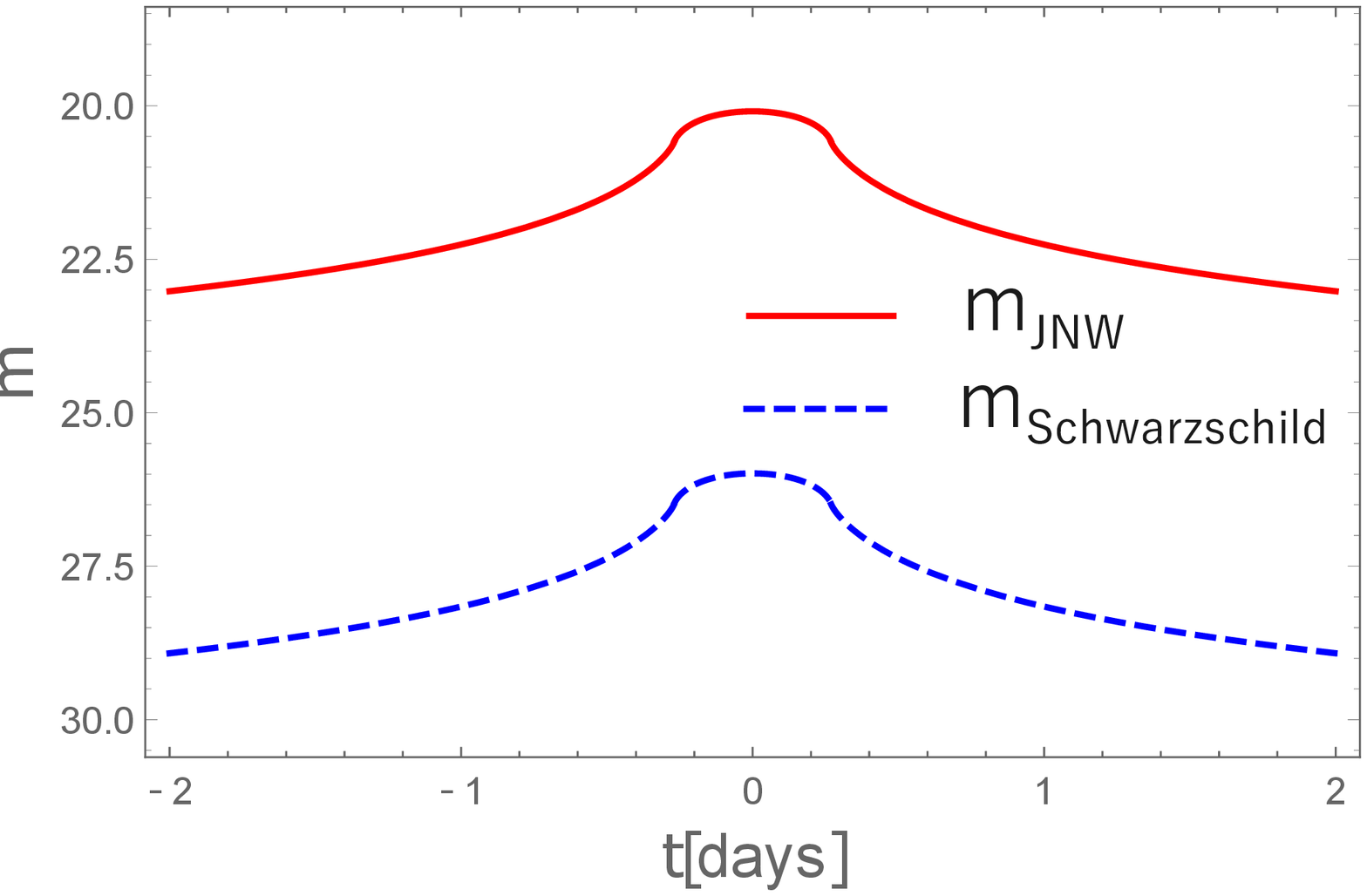}
    \includegraphics[scale=0.269]{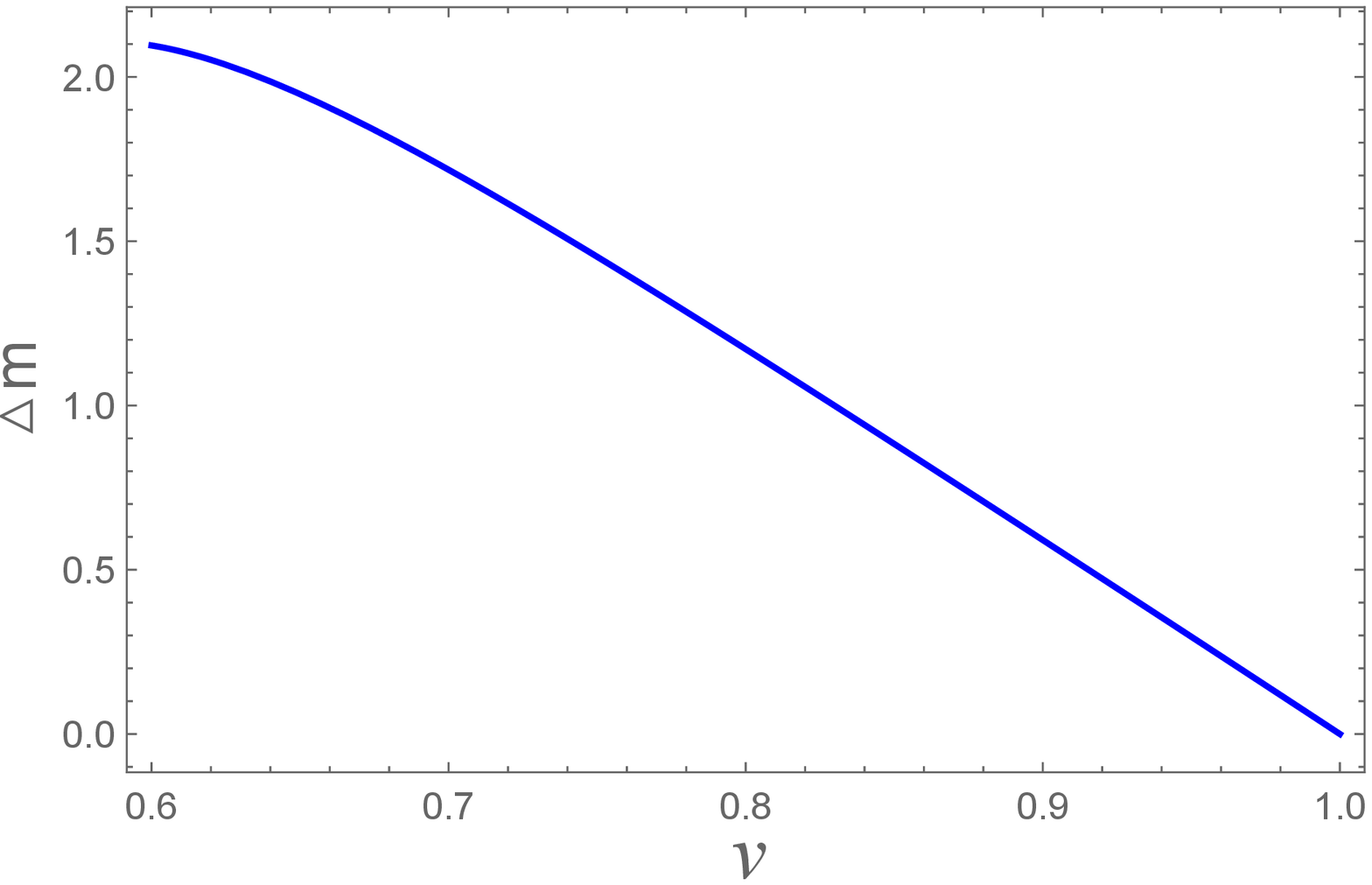}
     \includegraphics[scale=0.266]{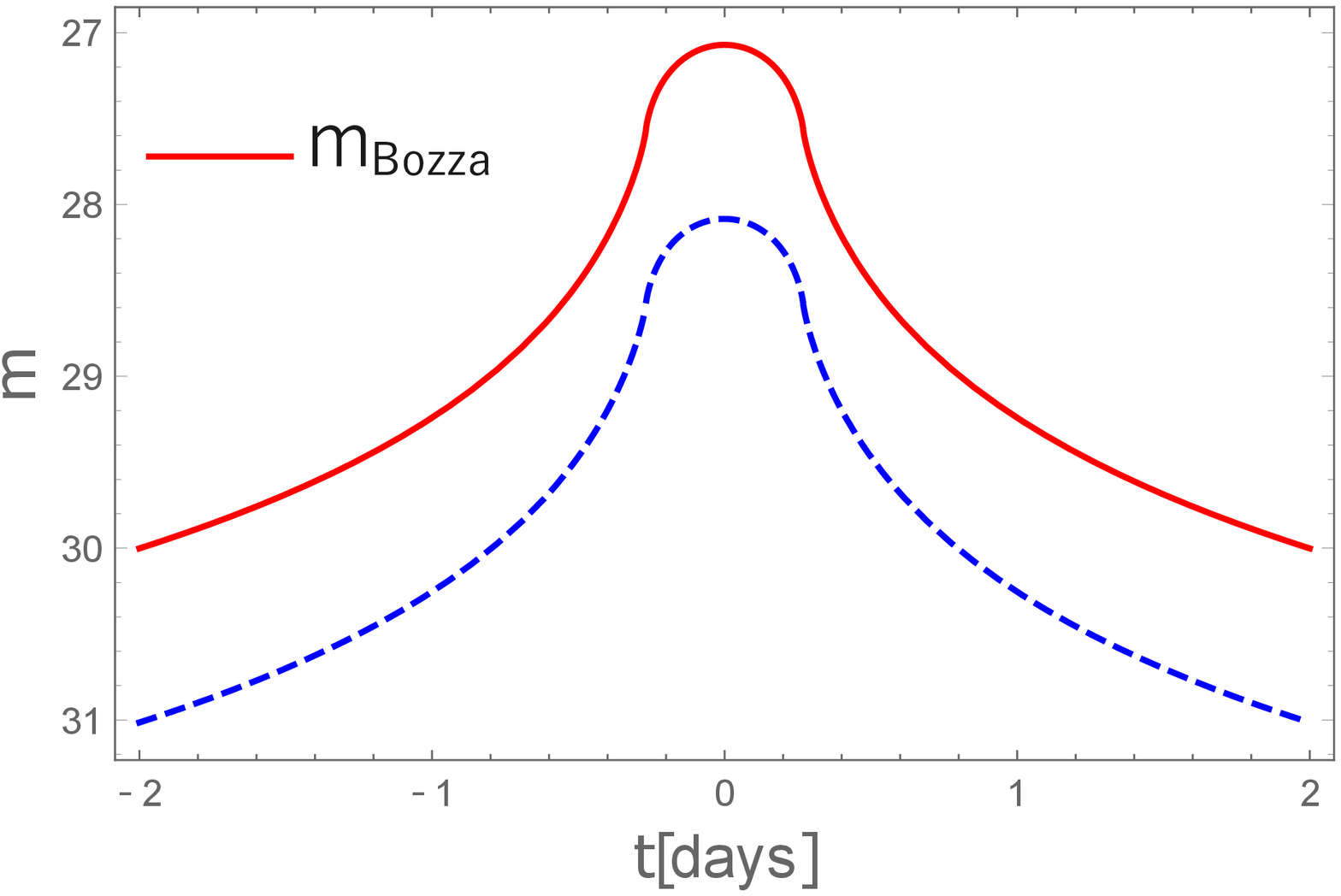}
  \end{center}
\caption{(Left Panel) The retrolensing light curves by a JNW naked singularity with the mass $M=10M_{\odot}$ at $D_{OL}=0.01$ pc for the scalar parameter: $\nu=0.51$ (red solid) and $\nu=1$ (blue dashed). (Middle Panel) The relative magnitude $\Delta\mathrm{m}$ dependence of the scalar parameter $\nu$. (Right panel) The apparent magnitude compared to \cite{Bozza:2002b} (red solid) in the presence of a strong scalar field.}\label{mnu}
\end{figure}

The light rays lensed by a gravitating body experience
a discernible change in its intrinsic spectrum, resulting in the magnification of the relativistic image luminosity. In order to investigate the evolution of image brightness by varying the scalar parameter $\nu$ we consider Sun as a source to study the light curves generally obtained from the apparent magnitude $\mathrm{m}$=$\mathrm{m}_{\odot}$-2.5$\log_{\mathrm{10}}(\mu_{\mathrm{tot}})$ of a variable star as a function of time \cite{Wheel:2002a,Fdepaol:2003a}.
Before proceeding any further, we adjust the position of the observer relative to the lens with mass $M=10M_{\odot}$ at $D_{OL}$=0.01pc.
In the left panel of Fig.~(\ref{mnu}), the image magnification illustrates a proportional relationship with the scalar field \cite{Vrbh:2008b}, simply put, the image governed by a naked singularity is more luminous in contrast to a black hole. As a matter of fact, the presence of scalar charge acts as a powerful complement to the gravitational force, which leads to maximum spectral distortion of the retrolensed light rays to create a magnified image. At $t=0$, referring to the peak of the light curve, we observe the relative magnitude $\Delta\mathrm{m}$ with respect to the gravity devoid of the scalar field by varying $\nu$, which unequivocally shows that $\mathrm{m}_\mathrm{JNW}>\mathrm{m}_\mathrm{Schwarzschild}$ (right panel). In a similar vein, we use the light curves to examine the difference between Ref. \cite{Bozza:2002b} and our results, thus, the visual representation of Bozza's equations for $\mathrm{m}$ produces comparatively higher results, see the middle panel.

\begin{figure*}[t]
\begin{center}
   \includegraphics[scale=0.35]{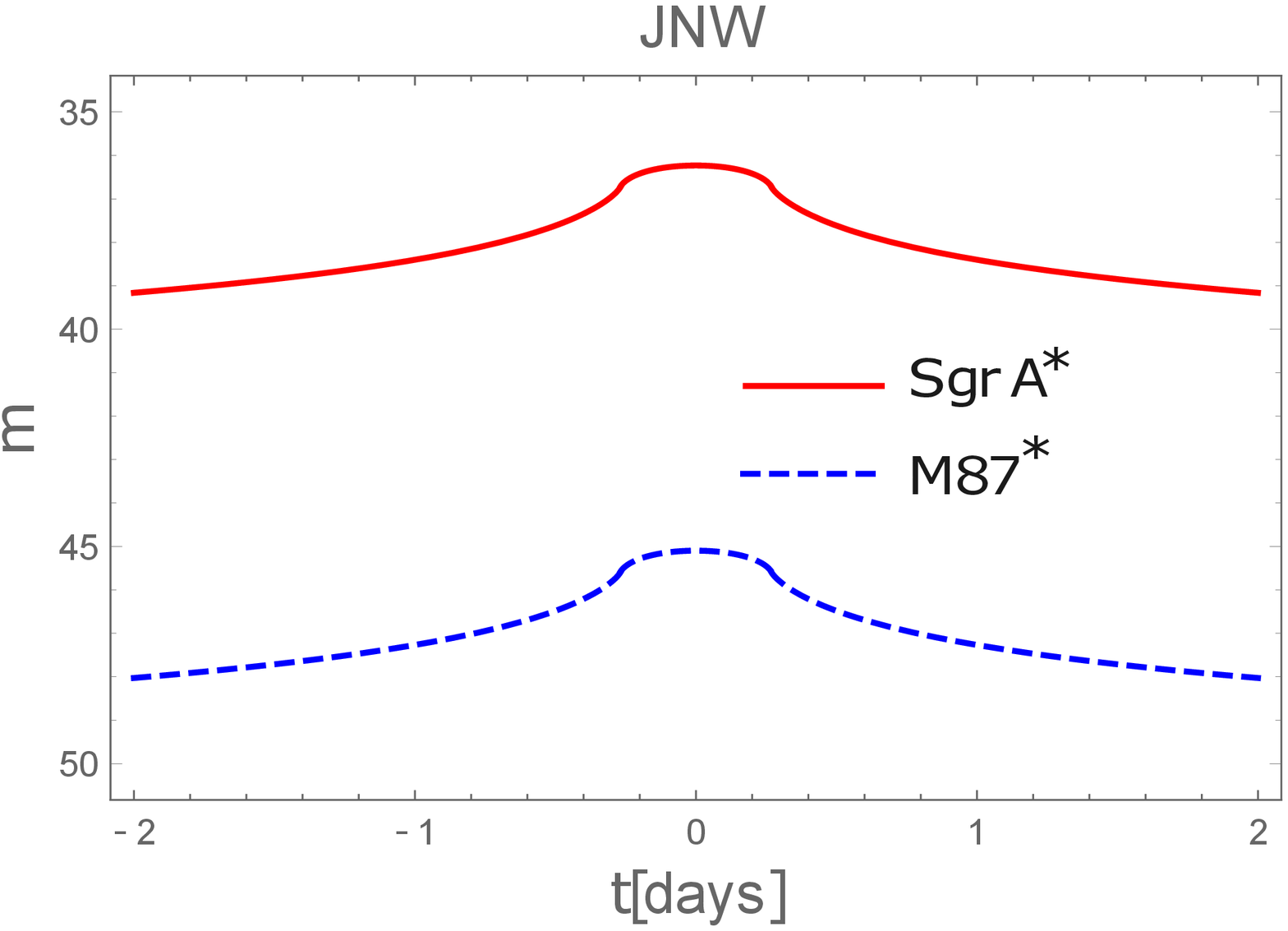}
   \includegraphics[scale=0.35]{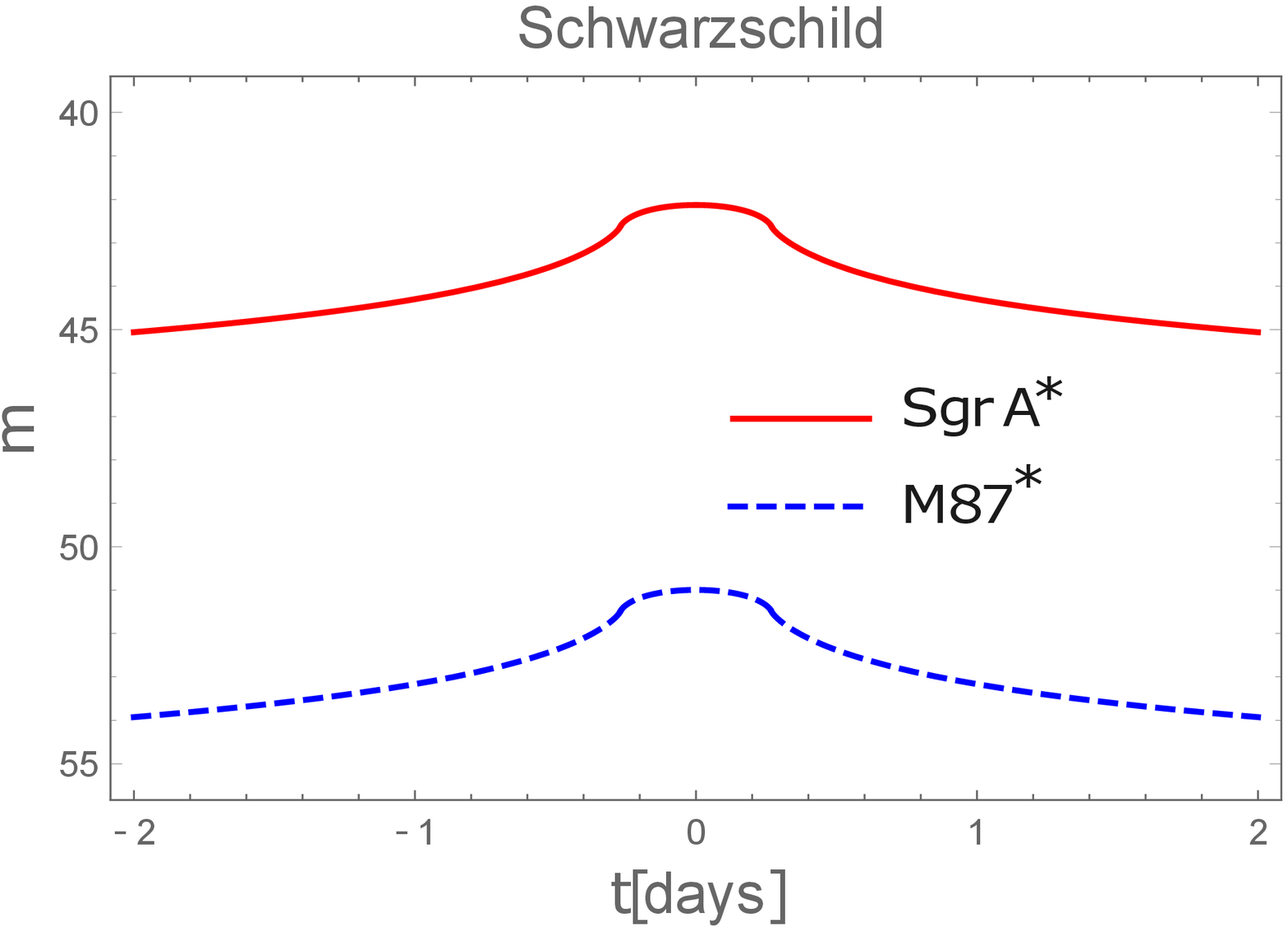}
  \end{center}
\caption{The retrolensing light curves for $\mathrm{Sgr} \mathrm{A}^*$ (red solid) and $\mathrm{M87}^*$ (blue dashed) acting as lens for a strong scalar gravity (left panel) and a non-scalar gravity (right panel).}\label{msgrM87}
\end{figure*}

For a better assessment of the JNW astrophysical properties we model the realistic nearby supermassive black holes, that is, Sgr A* and M87* as the lens, see Fig. (\ref{msgrM87}). In compliance with the literature the corresponding masses and distances from the Earth for Sgr A* \cite{Do:2019vob} are $M=4.31\pm0.38 \times 10^6 M_{\odot}$, $D_{OL}=7.94\pm0.42$ Kpc and for M87* \cite{Akiyama:2019vo} are $M = (6.5\pm 0.7) \times 10^9 M_{\odot}$, $D_{OL} = (16.8\pm0.8)$ Mpc. We only consider the mass and distance of the modeling lenses, while all other spacetime features remain intact.
Neglecting the error terms, the numerical values of maximum brightness for Sgr A* at $\nu$=0.51 and 1 are, respectively, recorded as
$\mathrm{m}$=-36.2324 and -42.1295 mag, likewise, M87* admits the values $\mathrm{m}$=-45.0979 and -50.9951 mag, see the left and right panel. The difference $\mathrm{m}_{\mathrm{SgrA^*}}-\mathrm{m}_{\mathrm{M87^*}}$=-8.865 mag remains invariant for each $\nu$. As expected, higher magnification is verified for a scalar field gravity.

\subsection{Retrolensing image}
Most commonly, a source behind a spherically symmetric black hole produces a countably infinite number of relativistic images in pairs, also, called the double image \cite{Bozza:2002b,Perlick:2002a}. This concept is posed with the deflection angle of light rays, i.e., $\pi-\alpha$ and $\pi+\alpha$, which, respectively, results in the appearance of a primary image and a secondary image on the opposite sides of the lens, centered on the source-observer plane \cite{Wheel:2002a}.
The separation $\theta_{+}-\theta_{-}$ of the double image is derived by means of
(\ref{theta+}) along with the term owing to the spherical symmetry,

\begin{align}\label{doubletheta}
\theta_{+}-\theta_{-}\sim2\theta_{+}=2\theta_{m}\bigg(1+\mathrm{exp}\bigg(\frac{\bar{b}-\pi+\beta}{\bar{a}}\bigg)\bigg).
\end{align}

\begin{figure*}[t]
 \begin{center}
   \includegraphics[scale=0.36]{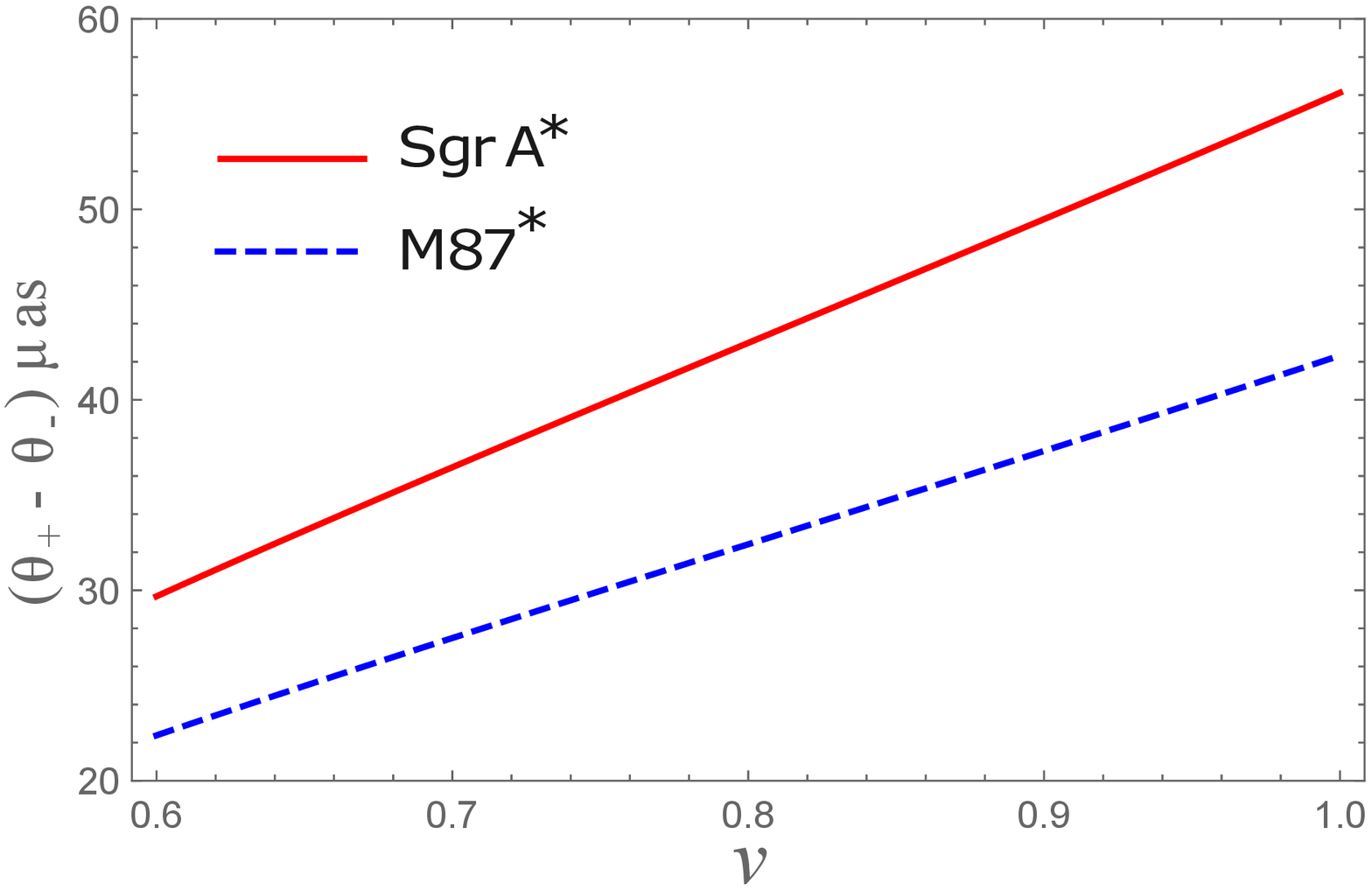}
   \includegraphics[scale=0.36]{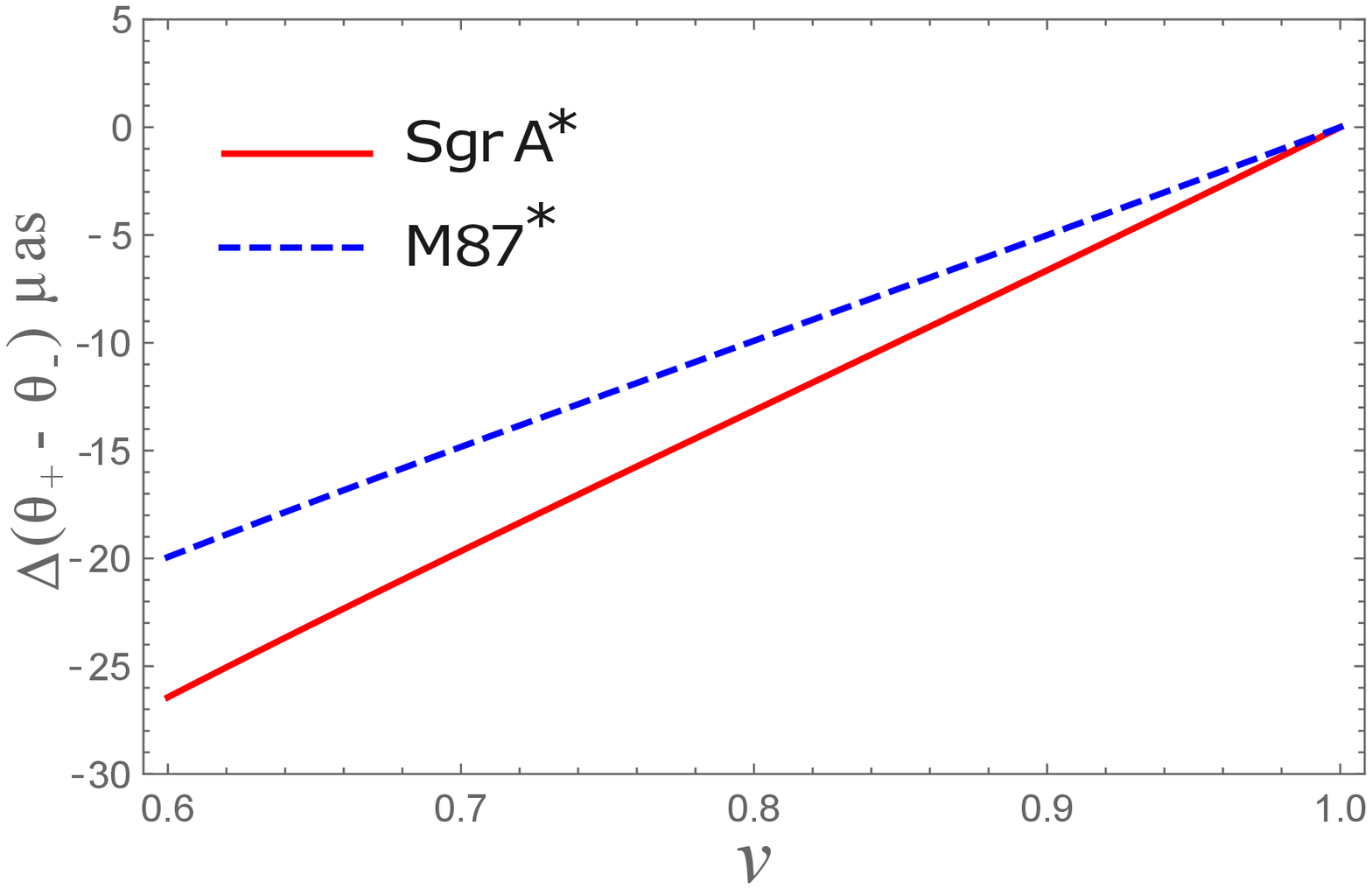}
  \end{center}
\caption{The separation $\theta_{+}-\theta_{-}$ (left panel) and the relative separation $\Delta(\theta_{+}-\theta_{-})$ (right panel) of a double image as a function of $\nu$ for Sgr A* (red solid) and M87* (blue dashed) acting as lens.}\label{doubletheta1}
\end{figure*}

Moreover, we define an expression $\Delta(\theta_{+}-\theta_{-})$, termed as the relative separation, which in the current context, determines the difference of the double image separation between a scalar and a non-scalar field gravity. Note that, $\theta$ in principle counts on the critical impact parameter $b_c$ since $e^{\frac{\bar{b}-\pi}{\bar{a}}}\ll$1, $\forall$ $\bar{a}>$0 and $\bar{b}<$0.
By taking Sgr A* and M87* as the lens frame, we analyse the behaviour of the angular position of the images and parity images (see, Fig.~(\ref{doubletheta1})), as well as, the effect of the lensing coefficients acknowledged in \cite{Bozza:2002b}. The attained results are summarized in Table (\ref{table6}). In general, the angular position experiences a decrease in case of a naked singularity and the values appears higher for \cite{Bozza:2002b}. According to \cite{Bozza:2002b}, based on the past data available regarding Sgr A*, the angular position was reported to be $\sim17$ $\mu\mathrm{as}$, whereas, the value reckoned from our calculations is  $\sim28$ $\mu\mathrm{as}$. Assuming Sgr A*, the average rate of change of $\theta_{+}$ and ${\theta_{+}}_{Bozza}$ is marked as 33.0543$\mu$as and 32.1233$\mu$as, respectively, for M87* each of these quantities turns out to be  24.925$\mu$as and 24.223$\mu$as. Here, the rate exceeds \cite{Bozza:2002b} due to the scalar dependent coefficients $\bar{a}$ and $\bar{b}$.

In the light of the above results, we can give an approximate estimation on the observability of $\theta_+$, $\theta_+-\theta_-$ and $\Delta(\theta_+-\theta_-)$. It is well known that the EHT collaboration \cite{aki:2019a} used a worldwide network of observatories, giving an effective observational aperture of $D\simeq 13\,400\;\mathrm{km}$ which is approximately the size of the Earth. Light from M87* was observed at $\lambda=1.3\;\mathrm{mm}$, resulting in an angular resolution of $\mathrm{Res}\simeq\frac{\lambda}{D}\simeq 20\;\mu\mathrm{as}$. This is within the range to observe $\theta_+-\theta_-$, but $\theta_+$ is feasible under the constraint $\frac{7}{10}<\nu<1$ for Sgr A* and $\frac{9}{10}<\nu<1$ for M87*.
On the other hand, an observation of $\Delta(\theta_+-\theta_-)$ still requires a future observational technology that can provide a higher angular resolution. However, space-based stations observatories may increase the resolution.
For instance, Ref.~\cite{Johnson:2019ljv} suggested that VLBI using lunar-based                                                                                                                                                       stations to resolve higher-order photon subrings. If such an observatory is available, roughly taking $D$ to be the radius of lunar orbit, (i.e., $D\simeq384\,400\;\mathrm{km}$), one finds $\mathrm{Res}\simeq 0.697\;\mu\mathrm{as}$, a great increase in resolution. (Our terminology `Earth-based' observer still applies here, as the Earth-Moon system is still approximately a point relative to the distances to the source and lens.)

\begin{table}[t]
\caption{The angular position, the angular separation and the relative angular separation of the parity images by taking Sgr A* and M87* as the lens for different values of the scalar parameter $\nu$. The unit used is microarcseconds ($\mu$as).\label{table6}}
\resizebox{1\textwidth}{!}{

	\begin{tabular}{p{0.5cm} p{1.4cm} p{1.4cm} p{1.4cm} p{2cm}p{1.8cm} p{1.4cm} p{1.4cm} p{1.4cm} p{2cm} p{1.8cm}}
		
\hline\hline
\multicolumn{1}{c}{}&
 \multicolumn{5}{c}{Sgr A*} & \multicolumn{5}{c}{M87*} \\
{$\nu$ } & {$\theta_{+}$} & {${\theta_{+}}_{Bozza}$} & {$\theta_{+}-\theta_{-}$} & {${\theta_{+}-\theta_{-}}_{Bozza}$}& {$\Delta$($\theta_{+}-\theta_{-}$)}& {$\theta_{+}$} & {${\theta_{+}}_{Bozza}$} & {$\theta_{+}-\theta_{-}$} & {${\theta_{+}-\theta_{-}}_{Bozza}$}& {$\Delta$($\theta_{+}-\theta_{-}$)}\\
\hline

0.6 & 14.8410 &15.2134   &29.6819  &30.4267    & -26.4435      & 11.1910   &11.4718     & 22.3820   &22.9436   & -19.9400       \\
0.7 &18.2297  &18.6055   &36.4594   &37.2110   &-19.6660       & 13.7463  &14.0300    & 27.4926   &28.0599   & -14.8294     \\
0.8 &21.4935  &21.8322   &42.9869  &43.6643    & -13.1385      & 16.2074  &16.4628    & 32.4147   &32.9255    & -9.90719     \\
0.9 &24.7433  &24.9731   &49.4865  &49.9462    & -6.63885      & 18.6579  &18.8312    & 37.3158   &37.6624    &-5.00609      \\
1   &28.0627  &28.0627   &56.1254   &56.1254   &0             & 21.1610   &21.1610    & 42.3219   &42.3219    &0              \\
\hline
	\end{tabular}
}	
\end{table}

\begin{figure}[t]
\begin{center}
   \includegraphics[scale=0.35]{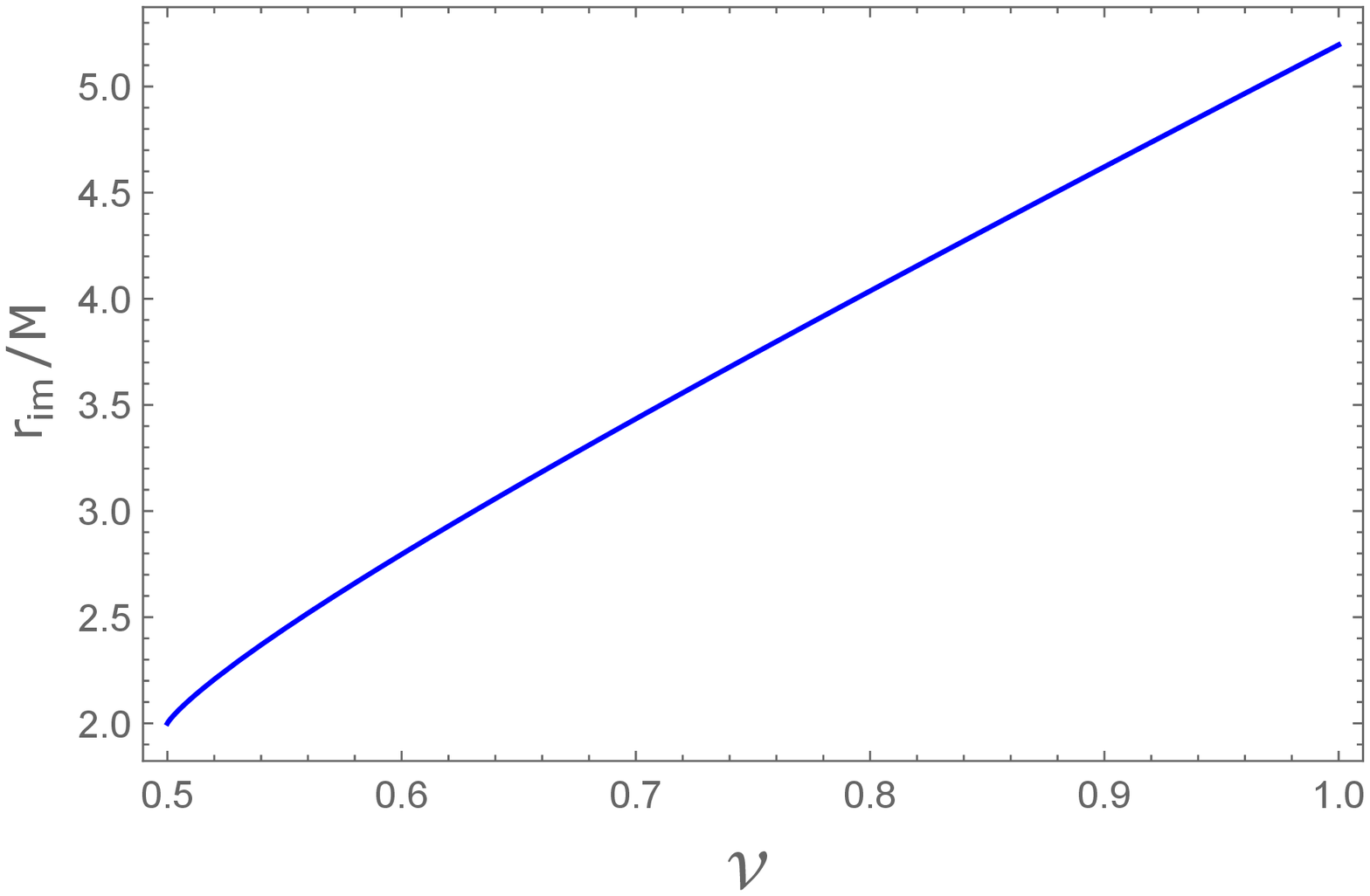}
   \includegraphics[scale=0.3]{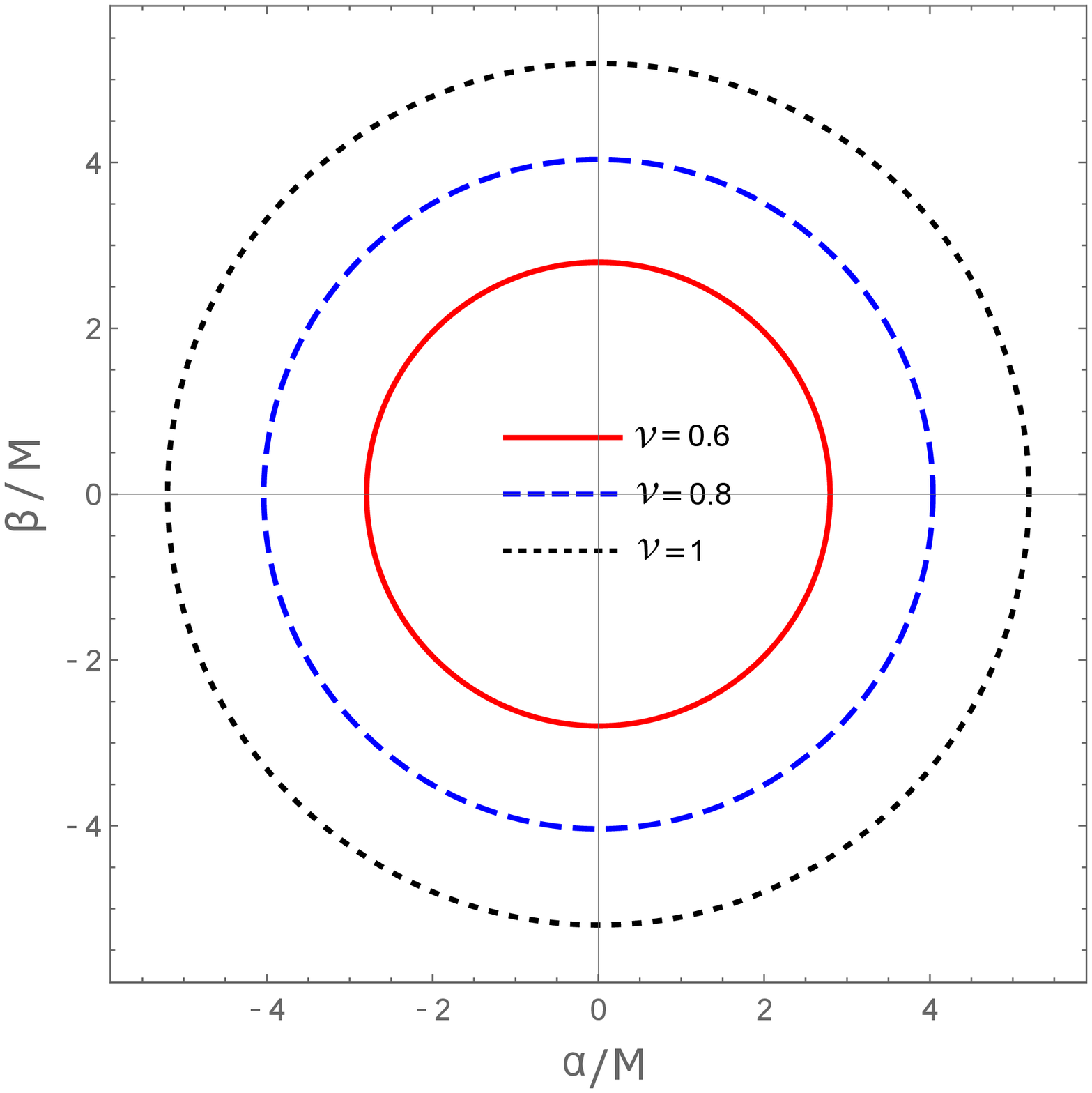}
  \end{center}
\caption{(Left panel) The radius $r_\mathrm{im}$ of the retroimage as a function of the scalar parameter $\nu$.
(Right panel) Ray-traced retroimages for $\nu$= 0.6 (red solid), 0.8 (blue dashed) and 1 (black dotted).}\label{Rsh}
\end{figure}

Now, we proceed with the classical Hamiltonian formulism to contemplate the contour of the image. The Hamilton-Jacobi equation for null geodesics associated with the particle's angular momenta $p_\mu$=$g_{\mu\sigma}\dot{x}^\mu$ is asserted by $H$=$\frac{1}{2}g_{\mu\sigma}p_{\mu}p_{\sigma}$=0. Subsequently, the photon trajectories are easily constructed out of the equations $\dot{x}^\mu$=$\frac{\partial H}{\partial p_\mu}$ and $\dot{p}_\mu$=$\frac{\partial H}{\partial x^\mu}$ $(\theta\neq\pi/2)$, as follows \cite{Babar:2017a}
\begin{align}
\dot{t}&=f^{-\nu}E, \quad \dot{\phi}=\frac{L}{r^2f^{1-\nu}\sin^2\theta}, \\
\dot{r}&=\sqrt{R}, \quad \dot{\theta}=\frac{\sqrt{\Theta}}{r^2f^{1-\nu}}.
\end{align}
The functions $R$ and $\Theta$ are presented as
\begin{align}
R&=E^2-\frac{(\mathcal{Q}+L^2)}{r^2f^{1-2\nu}}, \label{R}\\
\Theta&=\mathcal{Q}-L^2\cot^2\theta.
\end{align}
Here, $\mathcal{Q}$ is a conserved quantity known as the Carters's constant.
The image boundary is characterised by the celestial coordinates ($\alpha$, $\beta$) that are
obtained from the condition $R(r)$=$R'(r)$=0, inevitably fulfilled by an orbit with constant radius $r$ \cite{sub:2020a}.
Note that, $\alpha$ and $\beta$ are parameterized by the photon radius.

\begin{align}
\alpha^2+\beta^2=r_m^2f^{1-2\nu}.
\end{align}
Fig.~(\ref{Rsh}) displays the image radius $r_\mathrm{im}$ behaviour as a function of the scalar parameter; the stereographic projection of the celestial coordinates onto the 2D plane is presented in the right panel.
In contrary to a non-scalar field gravity the size shrinks down in the background of a scalar field gravity.
The curvature singularity is seen to appear immediately after the parameter $\nu$ reaches the critical value $\nu$=$\frac{1}{2}$, which is in agreement with
the findings of \cite{sub:2020a}. The Schwarzschild image radius 3$\sqrt{3}M$ restores at $\nu$=1 \cite{Babar:2020a}. In addition, by inserting $\nu$=1 in (\ref{theta+}) we compute 56 $\pm$8 and 42$\pm$3 as the corresponding angular diameter of the retroimages casted by Sgr A* and M87* \cite{cosmio:2013a,sub:2020a}.

\section{Conclusion} \label{con}
In this paper, we set up an investigation to explore the retrolensing phenomenon in a JNW spacetime considering a strong field paradigm.
We entirely embraced the notion of \cite{Tsuk:2017a} by introducing the variable $z$=$1-\frac{r_{0}}{r}$
and collated our results with \cite{Bozza:2002b}, where $z$=$\frac{-g_{tt}(r)+g_{tt}(r_0)}{1+g_{tt}(r_0)}$ functions as the variable.
In contrast to \cite{Bozza:2002b}, both the lensing coefficients are obtained in terms of the scalar parameter $\nu$.
In the course of our discussion, the value $\nu$=$\frac{1}{2}$ appeared as the critical value that triggers an unstable behaviour.
The realistic supermassive black holes Sgr A* and M87* are tooled to anticipate the behaviour of the light curves and the angular positions of the parity images. It turns out that the angular position and the angular separation of the parity images escalate for the black holes, on the other hand, the image magnification enhances in the case of naked singularities.
These results are also compatible with the comparison between a non-scalar and a scalar field gravity.
Taking into account all the outcomes in conformity with \cite{Bozza:2002b} and \cite{Tsuk:2017a},
we conclude that both the procedures well-define the physical features of the JNW naked singularity.
All in all, Tsukamoto's technique furnishes an analytical approach with a high precession to find the deflection angle in the strong field limit, moreover, it could be a useful stratagem to normalize the indeterminate physical problems in the future research.


\bibliographystyle{ieeetr}
\typeout{} \bibliography{Lensing}
\end{document}